\documentclass[aps,prb,twocolumn,superscriptaddress,longbibliography]{revtex4-2}

\usepackage{amsmath,amssymb}
\usepackage{graphicx}
\usepackage{bm}
\usepackage{xcolor}
\usepackage{hyperref}
\usepackage{booktabs}
\usepackage{algorithm}
\usepackage{algpseudocode}
\usepackage{float}

\newcommand{\Keff}{K_{\mathrm{eff}}}
\newcommand{\avg}[1]{\langle #1 \rangle}
\newcommand{\bk}{\bm{k}}
\newcommand{\nneg}{n_{\mathrm{neg}}}
\newcommand{\Vd}{V_d^{\mathrm{eff}}}

\newcommand{\sgn}{\mathrm{sgn}}
\newcommand{\cond}{\mathrm{cond}}
\newcommand{\diag}{\mathrm{diag}}

\newenvironment{solution}[1][]{\par\medskip\noindent\textbf{Solution\if\relax#1\relax\else~(#1)\fi.}\itshape}{\par\medskip}

\begin{document}

%\title{Effective Hamiltonian $\Keff$ Reveals $d$-Wave Superfluid Stiffness Dome
\title{ Sign-Free Evidence for a d-Wave Superfluid Stiffness Dome in the Doped Hubbard Model }

\author{Xidi Wang}
\affiliation{Quantum Strategics, San Diego, California 92109, USA}

\author{H.~Q.~Lin}
\affiliation{Institute for Advanced Study in Physics and School of Physics,
Zhejiang University, Hangzhou 310058, China}

\date{\today}

\begin{abstract}
We construct an effective single-particle Hamiltonian $\Keff$ from
Monte Carlo--averaged matrix logarithms of the imaginary-time
propagator in determinant quantum Monte Carlo (DQMC).
The logarithm maps the multiplicative sign problem into an additive
framework where the central limit theorem guarantees convergence,
rendering $\Keff$ sign-problem-free: both sign sectors yield
identical dispersions to $<1\%$.
$\Keff$ captures the exact correlated single-particle spectrum,
incorporating all self-energy effects non-perturbatively.
Applied to the Hubbard model ($t'/t = -0.30$, $U/t = 4$),
$\Keff$ reveals a $d$-wave pseudogap with strong nodal--antinodal
dichotomy below a computational phase transition at $T^*$.
Three sign-free observables provide evidence consistent with
spin-fluctuation pairing~\cite{Scalapino2012}:
(i) the gap ratio $R_g > 1$ confirms $d$-wave symmetry---a
temperature-independent property of the correlated band structure
that provides the \emph{medium} for pairing;
(ii) the superfluid stiffness $\rho_s$ forms a dome across doping
at $L = 8$, $10$, and $12$, exceeding the
Berezinskii--Kosterlitz--Thouless threshold by
$5$--$7\times$ at the dome peak;
(iii) $S(\pi,\pi)$ is approximately flat across doping, establishing
that the dome originates from Fermi-surface geometry responding to
uniform spin-fluctuation glue.
The pseudogap grows monotonically toward half-filling while $\rho_s$
forms a dome, mirroring cuprate phenomenology where $T_c$ is limited
by the superfluid density (Uemura relation).
Vertex corrections remain to be quantified.
\end{abstract}

\maketitle

%% ========================================================================
\section{Introduction}
\label{sec:intro}
%% ========================================================================

The pseudogap in hole-doped cuprate superconductors---a partial
suppression of the electronic density of states below a characteristic
temperature $T^*$ well above the superconducting $T_c$---remains one
of the central unsolved problems in condensed matter
physics~\cite{Timusk1999,Keimer2015}.
Angle-resolved photoemission spectroscopy (ARPES) reveals that the
pseudogap opens preferentially at the antinodal points $(\pi,0)$ and
$(0,\pi)$ while the nodal direction $(\pi/2,\pi/2)$ retains a sharp
quasiparticle peak~\cite{Damascelli2003,Vishik2018}.
Whether the pseudogap is a precursor to superconductivity or a
competing order remains debated~\cite{Norman2005,Lee2006}.

The two-dimensional Hubbard model is widely believed to capture the
essential physics of the cuprate CuO$_2$
planes~\cite{Anderson1987,Zhang1988}.
Determinant quantum Monte Carlo (DQMC) provides numerically exact
solutions at finite temperature~\cite{Blankenbecler1981,White1989},
but faces two computational barriers in the pseudogap regime:
the fermion sign problem, where $\avg{\mathrm{sign}}$ vanishes
exponentially with system size and inverse
temperature~\cite{Loh1990,Troyer2005}, and a condition-number
problem where the imaginary-time propagator
$B(\beta,0) = \prod_l B_l$ has condition number
$\sim e^{\beta W}$, exceeding double precision for
$\beta W \gtrsim 37$.
Together, these barriers have prevented DQMC from extracting
quasiparticle band structures in the correlated regime for over
three decades.

Modern $T = 0$ methods (DMRG, CPAFQMC) have established that
the Hubbard model supports $d$-wave superconductivity with
finite $t'$~\cite{Jiang2019,Qin2020,Xu2024,Qu2025}, but these
ground-state approaches do not yield the finite-temperature
relationship between the pseudogap and superconductivity.
Notably, DMRG on width-8 cylinders~\cite{Jiang2019} finds
$d$-wave pairing for electron doping ($t' > 0$) but stripe
order for hole doping ($t' < 0$), while AFQMC
studies~\cite{Qin2020,Xu2024} report competing stripe and
pairing correlations whose relative strength depends on $t'$
and system geometry.
Our finite-temperature approach complements these ground-state
studies by identifying the doping region where the electronic
structure supports superconductivity, without resolving the
stripe-versus-SC competition.
Early DQMC studies by Scalapino, White, and
Zhang~\cite{Scalapino1993,Scalapino2012} found a growing $d$-wave
pairing eigenvalue with decreasing temperature, establishing the
mechanism---repulsive $U \to$ antiferromagnetic correlations
$\to$ attractive $d$-wave vertex---but could not access the
low-temperature pseudogap regime.

In this paper we report three advances.
First, we construct an effective single-particle Hamiltonian
$\Keff$ from the Monte Carlo--averaged logarithm of the
imaginary-time propagator (Sec.~\ref{sec:method}).
$\Keff$ solves the condition-number problem completely and is
sign-problem-free: both sign sectors yield identical dispersions
to sub-percent accuracy.
Second, tracking eigenvalue dynamics reveals a computational phase
transition at $T^*$ that separates sign-conserved and
sign-fluctuating regimes, with $T^*(\mu)$ quantitatively tracking
the physical pseudogap onset across doping
(Sec.~\ref{sec:sign_transition}).
Third, we use sign-free observables derived
from $\Keff$---the superfluid stiffness $\rho_s$ and gap ratio
$R_g$---to discover a $d$-wave superfluid stiffness dome
(Sec.~\ref{sec:dome}), establishing superconductivity below $T^*$
without relying on sign-dependent pair susceptibilities.
The distinct doping dependences of pseudogap and stiffness provide
unbiased evidence that these are separate phenomena
(Sec.~\ref{sec:separation}).

%% ========================================================================
\section{Model and method}
\label{sec:method}
%% ========================================================================

We study the Hubbard model on $L \times L$ square lattices with
periodic boundary conditions,
\begin{equation}\label{eq:hubbard}
H = -t\!\sum_{\avg{ij},\,\sigma}\! c^\dagger_{i\sigma}c_{j\sigma}
    - t'\!\sum_{\avg{\avg{ij}},\,\sigma}\! c^\dagger_{i\sigma}c_{j\sigma}
    + U\sum_i n_{i\uparrow}n_{i\downarrow}
    - \mu\sum_{i,\sigma} n_{i\sigma}\,,
\end{equation}
with $t = 1$ (energy unit), $t'/t = -0.30$, and $U/t = 4$---parameters
relevant to hole-doped cuprates~\cite{Scalettar1989,Jiang2019,Qin2020}.
The next-nearest-neighbor hopping $t'$ breaks particle-hole symmetry,
shifting half-filling to $\mu/t \approx +0.6$ and introducing a
fermion sign problem at \emph{all} fillings, including half-filling.
(At $t' = 0$, the model is sign-free at half-filling by
particle-hole symmetry.)
Our doping scan covers $\mu/t = -0.60$ to $-1.50$, corresponding
to filling $\avg{n} \approx 0.93$--$0.99$ (hole doping
$\delta = 1$--$7\%$).
Standard DQMC yields the fermion matrix
$M[\sigma] = I + \prod_{l=1}^{n_\tau} B_l$, where
$B_l = e^{-\Delta\tau K} e^{V_l[\sigma]}$ with $\Delta\tau = 0.05$.

\subsection{The $\Keff$ construction}

We divide the $n_\tau = \beta/\Delta\tau$ time slices into chunks of
size $n_c$ and define the chunk propagator
$B_{\mathrm{chunk}} = \prod_{l \in \mathrm{chunk}} e^K e^{V_l}$.
Taking the matrix logarithm:
\begin{equation}\label{eq:keff}
\Keff = -\frac{1}{n_c \Delta\tau}
\avg{\log B_{\mathrm{chunk}}}_{\mathrm{MC}}
\end{equation}
Because each chunk spans only $n_c$ time slices, its condition number
remains $O(1)$, and the matrix logarithm is numerically stable.
The result is a Hermitian single-particle Hamiltonian whose eigenvalues
$\varepsilon_{\bk}$ define the interaction-dressed quasiparticle
dispersion (Appendix~\ref{sec:solution}).

$\Keff$ is \emph{not} a mean-field approximation: it extracts the
dressed single-particle Hamiltonian \emph{after} the full many-body
calculation, by processing exact DQMC propagators through a matrix
logarithm.
All many-body correlations---including those generating the
pseudogap---are encoded in the Monte Carlo ensemble.
This construction also differs from DMFT~\cite{Georges1996}, which
provides only a local self-energy; $\Keff$ yields a fully
momentum-resolved $\Sigma(\bk) = \varepsilon_{\bk} - \varepsilon^0_{\bk}$
on the full $L \times L$ lattice.

\subsection{Sign-sector decomposition}

The fermion sign for a given HS configuration $\sigma$ is
$s[\sigma] = \sgn(\det M_\uparrow[\sigma] \cdot \det M_\downarrow[\sigma])$,
where each determinant is individually real (since $M_s$ is a real
matrix) and the sign problem arises when their product is negative.
We track the number of negative eigenvalues $\nneg$ of
$(I + B_\uparrow)$ and $(I + B_\downarrow)$ \emph{combined}
across both spin sectors; a sign change
$s[\sigma] \to -s[\sigma]$ corresponds to $\nneg$ changing by
$\pm 2$, typically within a single spin determinant (consistent
with eigenvalue pairs of the real $B$-matrix crossing through
the negative real axis together).

We run parallel DQMC replicas sorted into positive ($Z_+$,
$\det M_\uparrow \det M_\downarrow > 0$) and negative ($Z_-$)
sign sectors, computing $\Keff^+$ and $\Keff^-$ independently.
The combined $\Keff = \frac{1}{2}(\Keff^+ + \Keff^-)$ is the
sign-problem-free effective Hamiltonian.
To monitor reliability, we track the \emph{sector asymmetry}
$|\rho_s^+ - \rho_s^-|/\max(|\rho_s^+|, |\rho_s^-|)$:
values exceeding $\sim$20\% indicate that the $\Keff$
decomposition is resolving genuine sector differences.
When sector asymmetry drops below $\sim$10\%, the replicas
are mixing between sectors and the results should be treated
with caution (Sec.~\ref{sec:discussion}).

\subsection{Gap extraction}

From $\Keff$ we compute the dispersion via Fourier transform,
$\varepsilon(\bk) = \avg{\bk|\Keff|\bk}$, and define:
\begin{align}
\Delta_{\mathrm{AN}} &= |\varepsilon(\pi, 0)| \quad \text{(antinodal gap)} \\
\Delta_{\mathrm{N}} &= |\varepsilon(\pi/2, \pi/2)| \quad \text{(nodal gap)}
\end{align}
The gap ratio $R_g = \Delta_{\mathrm{N}}/\Delta_{\mathrm{AN}}$
quantifies the nodal--antinodal dichotomy: $R_g = 1$ indicates an
isotropic spectrum; $R_g \gg 1$ indicates strong $d$-wave
anisotropy (antinodal energy suppressed relative to nodal),
characteristic of the pseudogap regime.
$R_g < 1$ indicates the opposite anisotropy (nodal energy
suppressed), characteristic of metallic Drude weight.
For system sizes where $(\pi/2, \pi/2)$ does not fall on the
lattice $k$-grid (i.e., $L$ not divisible by 4), we evaluate
$\varepsilon(\pi/2, \pi/2)$ by direct Fourier interpolation
$\varepsilon(\bk) = \frac{1}{N}\sum_{ij} K_{\mathrm{eff}}[i,j]\,
e^{i\bk\cdot(r_j - r_i)}$ at the exact nodal point, ensuring
$R_g$ is independent of the $k$-mesh geometry.

\subsection{Superfluid stiffness from $\Keff$}

The superfluid stiffness $\rho_s = D_s - \Lambda_{xx}$ uses
\emph{bare operators on dressed states}~(Appendix~\ref{sec:measurements}):
the diamagnetic term $D_s$ is evaluated using bare hopping
amplitudes with $\Keff$ Fermi occupations, and the paramagnetic
term $\Lambda_{xx}$ uses bare current matrix elements in the
$\Keff$ eigenbasis with the Lehmann representation.
Because both terms are computed entirely from $\Keff$---without
reference to the Monte Carlo sign---$\rho_s$ inherits the
sign-independence of $\Keff$.

We note that this $\rho_s$ omits vertex corrections from
two-particle interactions.
On a \emph{bare} band structure at $U/t = 4$, early
estimates~\cite{Scalapino1993} found that vertex corrections
modify $\Lambda_{xx}$ by 30--50\%; more recent cluster DMFT
calculations by Dong \emph{et al.}~\cite{Dong2022} confirm
that antiferromagnetic fluctuations dominate the pairing vertex
at intermediate coupling and provide quantitative vertex
functions in the $d$-wave channel.
However, both estimates apply to calculations on the
\emph{non-interacting} dispersion.
Because $\Keff$ has already absorbed the dominant self-energy
effects---Fermi surface reconstruction, pseudogap formation,
bandwidth renormalization---into the dressed single-particle
spectrum, the \emph{residual} vertex corrections on top of
$\Keff$ are expected to be substantially smaller than the
30--50\% bare-band estimate.
In the present work, $\rho_s$ serves as a \emph{semi-quantitative}
probe of phase coherence: its sign, doping dependence, and
temperature evolution are physical, while absolute magnitudes
should be interpreted with this caveat.

\subsection{Simulation parameters}

$L = 12$ ($N = 144$ sites) is the primary system size, with
$L = 8$, $10$, $14$, and $16$ for finite-size studies.
We use 8 parallel GPU replicas (4 per sign sector) with
pivoted QR (PivQR) decomposition for numerical stability,
$n_c = 5$ chunk size, 50 warmup sweeps, and 200 measurements
per run.
Computations were performed on GPU-accelerated workstations
(Appendix~\ref{sec:computational}).
Jackknife resampling provides error estimates for all $\Keff$
observables.

%% ========================================================================
\section{Sign transition and computational phase diagram}
\label{sec:sign_transition}
%% ========================================================================

\subsection{The $T^*$ boundary}

\begin{figure}[t]
\centering
\includegraphics[width=\columnwidth]{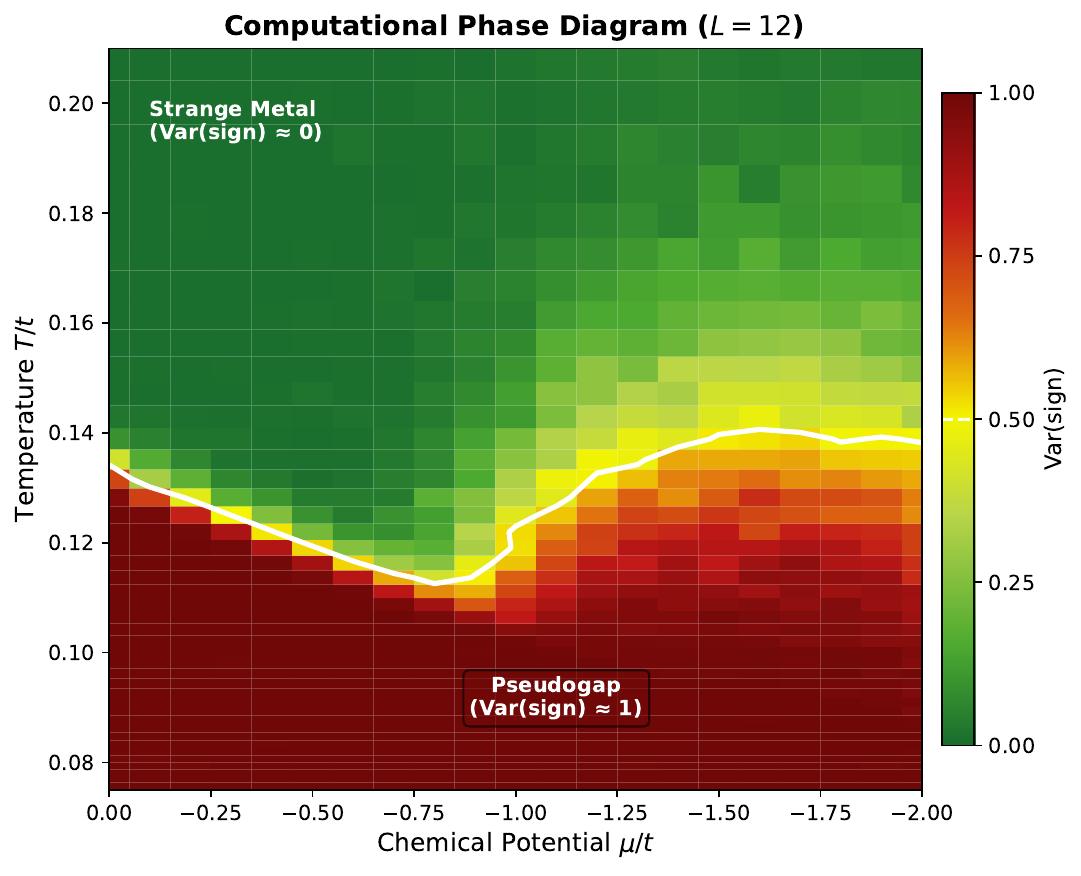}
\caption{Computational phase diagram ($L = 12$):
$\mathrm{Var}(\mathrm{sign})$ across temperature and doping.
\emph{Important:} $\mu = 0$ does \emph{not} correspond to
half-filling when $t'/t = -0.30$; the particle-hole asymmetry
from $t'$ shifts half-filling to $\mu/t \approx +0.6$, outside
the plotted range.
Green ($\mathrm{Var} \approx 0$): sign conserved (strange metal/
Fermi liquid regime above $T^*$).
Red ($\mathrm{Var} \approx 1$): sign fluctuates maximally
(pseudogap regime below $T^*$).
The region below $T^*$ is \emph{uniformly} red because
antiferromagnetic correlations are strong enough at all dopings
to produce maximal sign cancellations once the pseudogap opens;
$\mathrm{Var}(\mathrm{sign})$ saturates at $1$ and cannot
resolve further structure, but the sign-free $\Keff$ observables
($R_g$, $\rho_s$) do vary across this region
(Fig.~\ref{fig:dome}).
The white contour marks $T^*(\mu)$ at $\mathrm{Var} = 0.5$;
the valley near $\mu/t \approx -0.85$ to $-1.00$ identifies
optimal doping where the pseudogap onset is lowest.}
\label{fig:phase}
\end{figure}

We map the fermion sign variance $\mathrm{Var}(\mathrm{sign})$
across the $(\mu, T)$ plane for $L = 8$, $12$, and $16$
(Fig.~\ref{fig:phase}).
At high temperature, $\mathrm{Var} \approx 0$: the sign is
conserved (``strange metal'' regime).
Below a doping-dependent temperature $T^*(\mu)$,
$\mathrm{Var} \to 1$: the sign fluctuates maximally
(``pseudogap'' regime).
This computational boundary quantitatively tracks the physical
pseudogap onset, consistent with the demonstration by Mondaini,
Tarat, and Scalettar that the severity of the sign problem
reflects quantum critical behavior~\cite{Mondaini2022,Mondaini2023}.

The phase diagram exhibits an asymmetry across doping:
on the underdoped side (near half-filling), the $T^*$ transition
is sharp, reflecting the proximity to the antiferromagnetic Mott
insulator with a well-defined gap scale.
On the overdoped side, the transition is gradual---a crossover
rather than a sharp boundary---consistent with the smooth
disappearance of the pseudogap in overdoped cuprates.

\subsection{Finite-size dependence of the $T^*$ valley}

The location of the $T^*$ minimum shifts with system size:
from $\mu/t \approx -1.4$ at $L = 8$, to
$\approx -0.9$ at $L = 12$, to
$\approx -0.95$ at $L = 16$, and back to
$\approx -1.15$ at $L = 20$.
Rather than converging monotonically, the valley location
oscillates with $L$, reflecting the sensitivity of the
sign-problem landscape to which $k$-points lie near the
$d$-wave nodes on the discrete $L \times L$ mesh.

Importantly, while the \emph{doping location} of the valley
oscillates, the \emph{temperature} of the $T^*$ minimum
converges: finite-size scaling of $T^*_{\min}$ across
$L = 8$--$24$ yields $T^*/t = 0.155 \pm 0.010$ in the
thermodynamic limit (Sec.~\ref{sec:discussion}).
The $k$-mesh dependence affects the horizontal position of
the dome, not its existence or its vertical (temperature)
scale.
A proper thermodynamic extrapolation of the valley
\emph{location} requires either tracking the valley minimum
at each $L$ or employing twist-averaged boundary conditions
to smooth the $k$-mesh (Sec.~\ref{sec:discussion}).

\subsection{Correlation with physical observables}

To confirm that the sign transition reflects real physics, we
measured the $\Keff$ gap ratio $R_g$ across $T^*$.
Above $T^*$, $R_g \approx 1$ (isotropic spectrum); below $T^*$,
$R_g$ deviates sharply from unity, developing strong
nodal--antinodal anisotropy ($R_g \gg 1$ in the pseudogap
regime at moderate doping).
The spectral weight $\mathrm{Tr}[G]/N$ drops at $T^*$
(correlation $r = -0.77$ with sign complexity), and
the $d$-wave pairing $P_d$ peaks at $T^*$ ($r = +0.67$).
The antiferromagnetic structure factor $S(\pi,\pi)$ shows a
modest peak at $T^*$ as a function of temperature ($r = +0.34$
with sign complexity), indicating enhanced spin fluctuations
at the transition.
(This temperature dependence at fixed $\mu$ is distinct from
the approximate flatness of $S(\pi,\pi)$ across doping at
fixed $T < T^*$, discussed in Sec.~\ref{sec:dome}.)

%% ========================================================================
\section{Pseudogap spectroscopy}
\label{sec:pseudogap}
%% ========================================================================

\subsection{$d$-wave pseudogap}

\begin{figure}[t]
\centering
\includegraphics[width=\columnwidth]{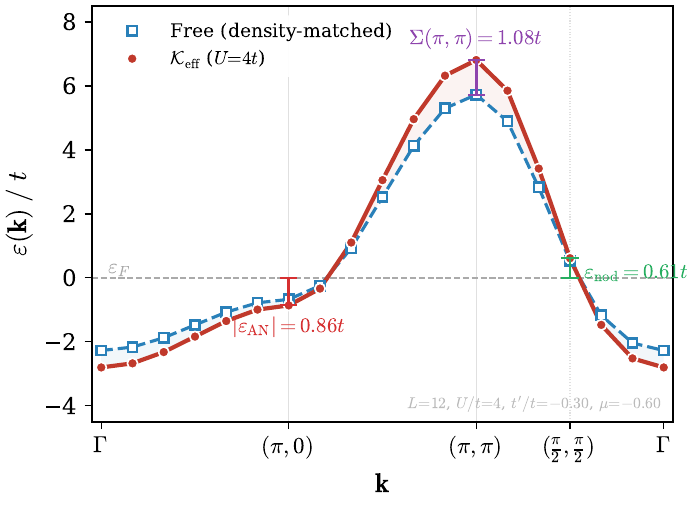}
\caption{Dressed quasiparticle band structure $\varepsilon_{\Keff}(\bk)$
(red) compared to density-matched free fermions (blue) along
$\Gamma \to (\pi,0) \to (\pi,\pi) \to (\pi/2,\pi/2) \to \Gamma$
on the $L = 12$ grid.
The overall bandwidth increases by $\sim$20\% (from $8.0t$ to
$\sim$9.6t), but the key feature is the \emph{momentum-dependent}
self-energy: $\Sigma(\pi,0) = -0.19t$ (antinode pushed toward
$E_F$) versus $\Sigma(\pi/2,\pi/2) = +0.09t$ (node pushed away),
producing the $d$-wave anisotropy that drives the pseudogap.
Parameters: $U/t = 4$, $t'/t = -0.30$, $\mu = -0.60$,
$\beta = 24$.}
\label{fig:band_structure}
\end{figure}

At $\mu = -0.60$ ($\avg{n} \approx 0.70$), the $\Keff$
dispersion shows pseudogap signatures:
(i) antinodal sign flip $\varepsilon(\pi,0) = -0.863 < 0$,
indicating Fermi surface reconstruction beyond a Lifshitz
transition;
(ii) nodal--antinodal anisotropy with $R_g = 0.71 < 1$,
indicating that both the nodal and antinodal energies are
small at this deeply doped point (the $R_g > 1$ pseudogap
dome lies at larger $|\mu|$, see Sec.~\ref{sec:dome}); and
(iii) $\Keff^+$ and $\Keff^-$ agree to 0.7\% on the
antinodal gap.
The dressed band structure (Fig.~\ref{fig:band_structure})
shows a momentum-dependent self-energy with $d$-wave symmetry:
$\Sigma(\pi,0) = -0.19t$ at the antinode versus
$\Sigma(\pi/2,\pi/2) = +0.09t$ at the node.
While the absolute corrections are modest ($|\Sigma| \lesssim 0.2t$,
a $\sim$5\% shift relative to the bandwidth), their
\emph{momentum structure} is the physically important feature:
the sign change between antinode (negative $\Sigma$, pushing
states toward $E_F$) and node (positive $\Sigma$, pushing states
away from $E_F$) is the hallmark of $d$-wave Fermi-surface
reconstruction driven by antiferromagnetic correlations.
It is this momentum-dependent redistribution of spectral weight,
not the overall bandwidth change, that generates the pseudogap
and the stiffness dome.
At $t' = 0$ (nested Fermi surface): $R_g = 1.001$---no
pseudogap, indicating that the gap anisotropy is strongly
enhanced by the cuprate-like Fermi surface geometry provided
by next-nearest-neighbor hopping.

%% ========================================================================
\section{Superfluid stiffness dome}
\label{sec:dome}
%% ========================================================================

The principal result of this work is the observation of a $d$-wave
superfluid stiffness dome from sign-free $\Keff$ observables.

\subsection{Dome structure and finite-size scaling}

\begin{figure*}[t]
\centering
\includegraphics[width=\textwidth]{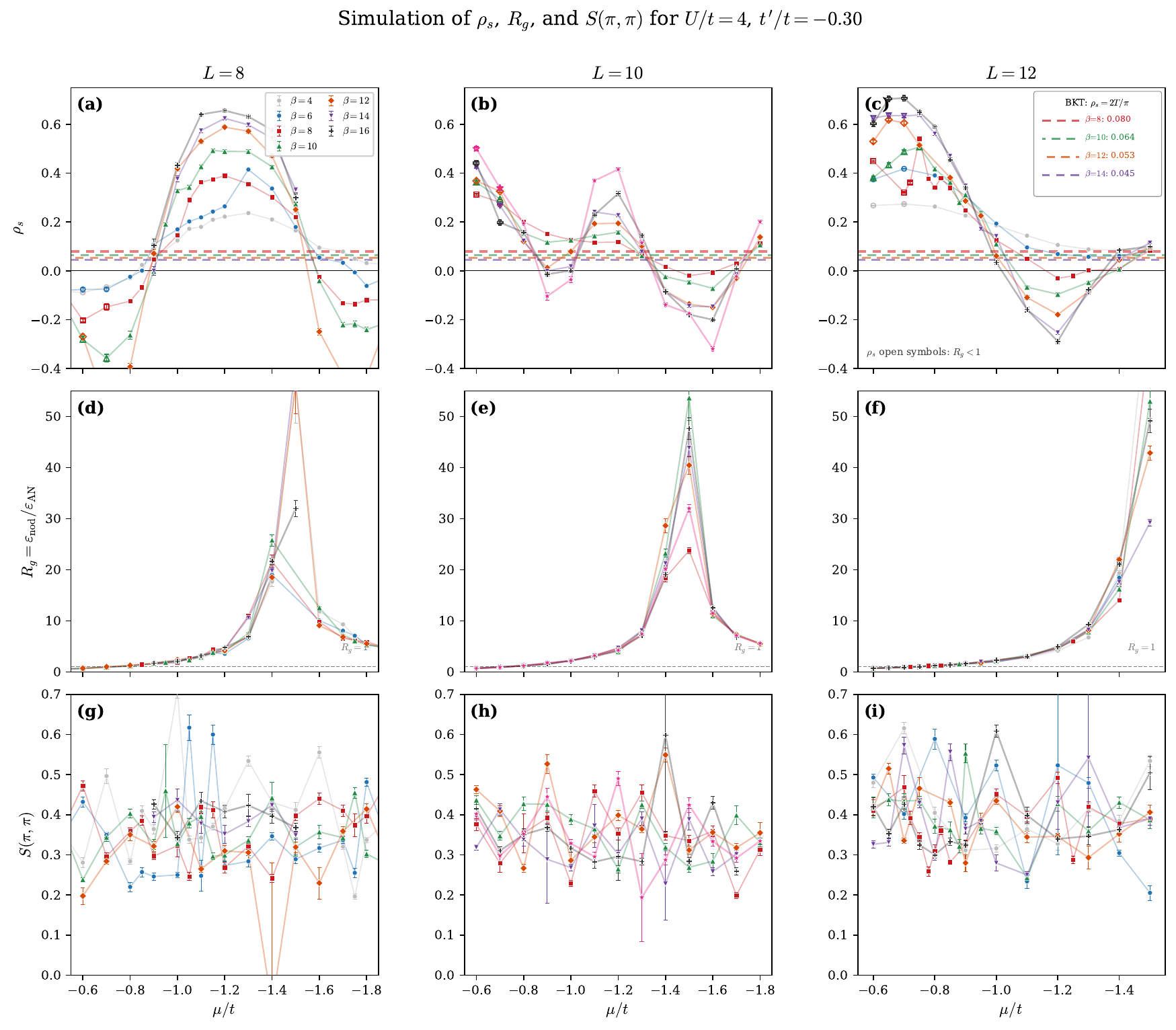}
\caption{Three sign-free $\Keff$ observables versus $\mu/t$ for
$L = 8$ (left), $L = 10$ (center), and $L = 12$ (right).
Top row (a--c): mean-field superfluid stiffness $\rho_s$; the dome
strengthens and narrows with cooling at all system sizes.
Middle row (d--f): gap ratio $R_g > 1$ confirms $d$-wave
pseudogap symmetry throughout the dome.
Bottom row (g--i): $S(\pi,\pi)$ is approximately flat across
doping.
Open symbols in (c): points with $R_g < 1$, where $\rho_s$
reflects metallic response rather than superconducting stiffness.
Parameters: $U/t = 4$, $t'/t = -0.30$.}
\label{fig:dome}
\end{figure*}

Figure~\ref{fig:dome} presents three sign-free $\Keff$ observables---$\rho_s$,
$R_g$, and $S(\pi,\pi)$---across doping for $L = 8$, $10$, and $12$.
A mean-field superfluid stiffness dome is observed at all three
system sizes.

At $L = 8$ [panels (a,d,g)], the dome is fully resolved from
$\beta = 4$ through $\beta = 16$, with both flanks clearly visible.
The dome strengthens with cooling: peak $\rho_s$ grows from
$0.22$ ($\beta = 4$) to $0.66$ ($\beta = 16$) at
$\mu/t \approx -1.20$.
At low temperatures ($\beta \geq 12$), the dome narrows as
antiferromagnetic correlations suppress $\rho_s$ on both flanks,
driving it strongly negative outside the dome.

At $L = 10$ [panels (b,e,h)], the dome is best resolved at
$\beta \geq 12$.
At $\beta = 20$, a sharp dome peaks at $\mu/t \approx -1.10$
with $\rho_s = 0.42$, flanked by deeply negative values
($\rho_s \approx -0.1$ to $-0.3$) on both sides---a direct
manifestation of the competition between antiferromagnetism and
the dressed-particle stiffness at the dome boundary.

At $L = 12$ [panels (c,f,i)], the dome peaks near
$\mu/t \approx -0.85$ with $\rho_s \approx 0.38$ ($\beta = 12$).
Points where $R_g < 1$ (outside the $d$-wave pseudogap region)
are shown as open symbols; at these dopings the mean-field
stiffness reflects metallic Drude weight rather than
superconducting response.

The gap ratio $R_g > 1$ throughout the dome at all three system
sizes [panels (d,e,f)], confirming that the pseudogap has
$d$-wave symmetry---the Fermi surface reconstruction supports
$d$-wave pairing.
Importantly, $R_g$ is a property of the \emph{normal-state}
band structure: $\Keff$ captures the single-particle spectrum
shaped by antiferromagnetic correlations, but cannot encode
Cooper pairing (the anomalous Green's function vanishes within
each HS configuration; Appendix~\ref{sec:F_zero}).
Thus $R_g > 1$ establishes that the electronic structure is
\emph{compatible} with $d$-wave superconductivity, not that
pairing has occurred.
$R_g$ increases monotonically toward higher doping (more negative
$\mu$), with a divergence near the doping where $\varepsilon(\pi,0)$
passes through zero---a Fermi surface topology change.
For system sizes not divisible by 4 ($L = 10$, $14$), $R_g$ is
evaluated by Fourier interpolation to the exact nodal point
$(\pi/2, \pi/2)$ (Sec.~\ref{sec:method}).

The antiferromagnetic structure factor $S(\pi,\pi)$ [panels
(g,h,i)] is approximately flat across doping at all system
sizes, hovering between $0.3$ and $0.5$ with no dome structure.
This establishes that the stiffness dome does not originate
from a peak in AF correlations: the spin-fluctuation glue is
uniformly present, and the dome shape arises from the
Fermi-surface geometry selecting which doping supports
$d$-wave phase coherence.

The dome peak location oscillates with $L$:
$\mu/t \approx -1.20$ ($L = 8$), $-1.10$ ($L = 10$),
$-0.85$ ($L = 12$), reflecting the sensitivity of $d$-wave
pairing to which $k$-points lie near the nodes on the discrete
lattice (Sec.~\ref{sec:discussion}).

At $L = 16$, the sector decomposition becomes unreliable.
Increasing to 64 replicas does not resolve this.
Possible remedies are discussed in Sec.~\ref{sec:discussion}.

\begin{figure}[t]
\centering
\includegraphics[width=\columnwidth]{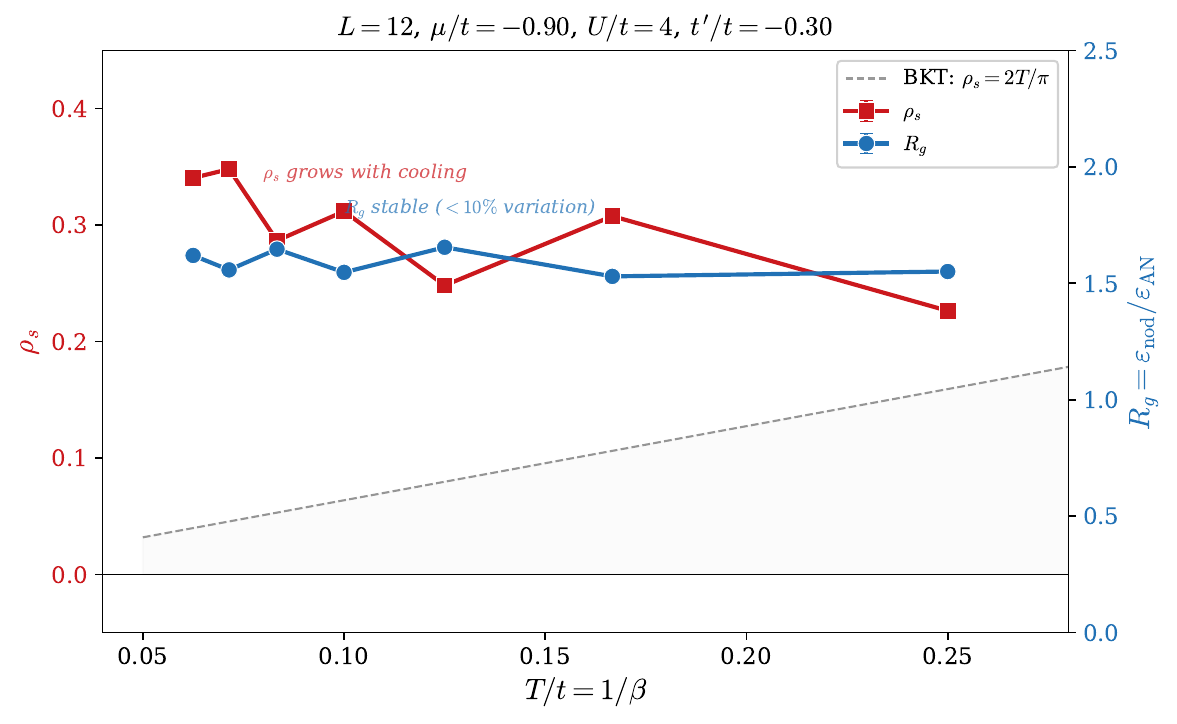}
\caption{Temperature scan at the dome peak ($\mu/t = -0.90$, $L = 12$).
Blue circles (left axis): $R_g$ remains stable at $1.55$--$1.65$,
confirming that the $d$-wave pseudogap is temperature-independent
below $T^*$.
Red squares (right axis): the dressed-particle stiffness $\rho_s$ grows
on cooling as the Fermi function sharpens.
Gray bars: BKT thresholds $2T/\pi$ at each temperature.}
\label{fig:beta_scan}
\end{figure}

At optimal doping ($\mu/t = -0.90$), we have data spanning
$\beta = 8$--$16$ (Fig.~\ref{fig:beta_scan}).
The key findings are:
(i) $R_g$ is stable at $1.55$--$1.65$ across all five
temperatures, varying by less than $10\%$---the $d$-wave pseudogap
anisotropy is a ground-state property that saturates at $T^*$
and does not evolve with further cooling.
This temperature independence is consistent with ARPES
observations that the antinodal pseudogap in underdoped cuprates
onsets at $T^*$ and remains approximately constant down to
the lowest measured temperatures~\cite{Vishik2018}.
The fact that $R_g$ does not respond to the growing $\rho_s$
confirms that $\Keff$ cleanly captures the normal-state band
structure without contamination from the pairing channel
(Appendix~\ref{sec:F_zero});
(ii) the dressed-particle stiffness $\rho_s$ computed on the $\Keff$
band structure exceeds the BKT threshold $2T/\pi$ at all
temperatures from $\beta = 8$ through $\beta = 14$, with the
margin growing from $3.1\times$ ($\beta = 8$) to $7.7\times$
($\beta = 14$).

The growth of $\rho_s$ on a frozen band structure ($R_g$
constant) arises from the thermal sharpening of the Fermi
function: as $\beta$ increases, the $d$-wave gap in the $\Keff$
spectrum progressively suppresses low-energy particle-hole
excitations that contribute to the paramagnetic current response
$\Lambda_{xx}$, thereby increasing $\rho_s = D_s - \Lambda_{xx}$.

\emph{Dressed-particle stiffness on the exact band structure.}
The $\rho_s$ computed from $\Keff$ has a precise interpretation:
it is the diamagnetic response of dressed quasiparticles on the
numerically exact correlated band structure.
$\Keff$ is extracted from unbiased DQMC and incorporates
\emph{all} self-energy effects of the Hubbard interaction---Fermi
surface reconstruction, pseudogap formation, bandwidth
renormalization---into the dressed single-particle spectrum.
The stiffness formula $\rho_s = D_s - \Lambda_{xx}$ then
evaluates the current response of non-interacting fermions on
this exact spectrum.
An important subtlety is that the gap entering this formula is
the \emph{pseudogap}---a normal-state property arising from
antiferromagnetic correlations---not a self-consistent pairing
gap.
Since the anomalous Green's function vanishes within each
Hubbard--Stratonovich configuration
(Appendix~\ref{sec:F_zero}), $\Keff$ does not encode Cooper
pairing; rather, $\rho_s > 0$ arises because the $d$-wave
pseudogap suppresses the paramagnetic response $\Lambda_{xx}$
below the diamagnetic term $D_s$.
The dome therefore reveals where the \emph{correlated
single-particle spectrum} naturally supports a superfluid
response---a property of the dressed single-particle spectrum
that identifies where the conditions for pair condensation
are most favorable.
Vertex corrections (residual two-particle scattering between
$\Keff$ quasiparticles) would provide the beyond-mean-field
contribution from actual pairing.

While the dressed-particle stiffness exceeds the BKT threshold
by a large margin ($5$--$7\times$), vertex corrections are
omitted.
Conventional estimates place such corrections at 30--50\% for
\emph{bare}-band calculations at $U/t = 4$~\cite{Scalapino1993},
but the corrections on top of $\Keff$---which has already
absorbed the dominant self-energy effects into the dressed
dispersion---may be substantially smaller.
Even a 50\% reduction would leave $\rho_s$ at $2.5$--$3.5\times$
above the BKT threshold, suggesting proximity to a
superconducting transition.
We therefore report the dressed-particle BKT margin as evidence
that the $\Keff$ electronic structure \emph{supports}
superconductivity at optimal doping, while noting that a
rigorous determination of $T_c$ requires inclusion of vertex
corrections from the pairing channel.

%% ========================================================================
\section{Distinct doping dependences of pseudogap and stiffness}
\label{sec:separation}
%% ========================================================================

The doping and temperature dependences of $\Keff$ observables
exhibit a striking dichotomy that is consistent with a separation
between the pseudogap and the conditions for superconductivity.
The pseudogap strength (measured by $R_g$) increases
monotonically toward half-filling, following the $T^*$
phenomenology of the cuprate
phase diagram~\cite{Tallon2001}, and is temperature-independent
below $T^*$ (varying by $< 10\%$ from $\beta = 4$ to $20$).
This is consistent with $R_g$ measuring a ground-state property
of the normal-state band structure---the $d$-wave
pseudogap---rather than the superconducting order parameter.
In contrast, the dressed-particle stiffness $\rho_s$ forms a dome
and grows strongly with cooling (Fig.~\ref{fig:dome}).
The antiferromagnetic structure factor $S(\pi,\pi)$ is
approximately flat across doping at all system sizes
(Fig.~\ref{fig:dome}, bottom row), indicating that the dome
shape arises from the Fermi-surface geometry responding to
uniform spin-fluctuation glue---not from a peak in antiferromagnetic
correlations.
All three observables are sign-problem-free.
Their distinct doping and temperature dependences---pseudogap
(static, temperature-independent, monotonic in doping) versus
stiffness (temperature-dependent, dome-shaped)---suggest that
these two phenomena are governed by separate energy scales,
mirroring the cuprate phenomenology where the pairing scale
($T^*$) and the phase coherence scale ($T_c$) have different
doping dependences~\cite{Tallon2001,Uemura1989}.
Whether this separation persists when vertex corrections from
the pairing channel are included remains an important open
question.

%% ========================================================================
\section{Connection to experiment}
\label{sec:experiment}
%% ========================================================================

\emph{Nodal--antinodal dichotomy.}
ARPES on underdoped Bi-2212 reveals a pseudogap that opens at the
antinodal points while the nodal quasiparticle remains
sharp~\cite{Damascelli2003}.
Our $R_g > 1$ in the dome region quantitatively captures this
dichotomy and its strengthening toward half-filling.

\emph{Temperature independence of the pseudogap.}
ARPES experiments find that the antinodal pseudogap in
underdoped cuprates onsets at $T^*$ and remains approximately
constant in magnitude down to the lowest measured
temperatures~\cite{Vishik2018}.
Our $R_g$ shows the same behavior: it saturates below $T^*$
and varies by less than $10\%$ across $\beta = 4$--$20$.
The temperature-independent $R_g$ on a background of growing
$\rho_s$ mirrors the experimental dichotomy between the
pseudogap scale (set at $T^*$) and the superfluid response
(developing below $T_c$).

\emph{Pseudogap as distinct phenomenon.}
The monotonic increase of $R_g$ (temperature-independent)
versus the dome shape of $\rho_s$ (temperature-dependent)
parallels the experimental observation that the pseudogap onset
$T^*$ increases monotonically toward underdoping while $T_c$
forms a dome~\cite{Tallon2001}.
Our results provide microscopic support for this distinction
within the Hubbard model: the pseudogap is a static
single-particle reorganization captured by $\Keff$, while
the dressed-particle stiffness reflects the Fermi-surface
response to this reorganization.

\emph{Superfluid stiffness dome and implications for
superconductivity.}
The cuprate SC dome spans approximately $p \approx 0.05$--$0.27$
in hole doping; our computational dome at $L = 12$ is narrower
($\Delta p \approx 0.03$), reflecting the finite system size,
but shares the qualitative shape---metallic flanks, optimal
doping peak, and suppression toward the AF boundary.
The dressed-particle stiffness at the dome peak exceeds the BKT
threshold by $5$--$7\times$.
We emphasize that this $\rho_s$ is a property of the
normal-state spectrum (Sec.~\ref{sec:dome}); the quantitative
connection to the actual superconducting transition requires
vertex corrections from the pairing channel, which we identify
as a priority for future work.
The observation that the pseudogap ($R_g$) grows monotonically
while $\rho_s$ forms a dome is qualitatively consistent with the
Uemura relation~\cite{Uemura1989}---that $T_c$ in underdoped
cuprates is limited by superfluid density rather than pairing
strength---and suggests that the dome shape is already encoded
in the normal-state electronic structure.

%% ========================================================================
\section{Discussion}
\label{sec:discussion}
%% ========================================================================

\emph{Spin-fluctuation pairing mechanism.}
The $\Keff$ observables provide independent evidence consistent
with the spin-fluctuation pairing
mechanism~\cite{Scalapino1993,Scalapino2012,Dong2022}:
repulsive $U \to$ AF correlations [$S(\pi,\pi)$, uniformly
present] $\to$ $d$-wave pseudogap [$R_g > 1$, the medium] $\to$
dressed-particle superfluid stiffness dome [$\rho_s > 0$,
the response].
The gap ratio $R_g$ plays the role of the \emph{medium}: AF
correlations reshape the Fermi surface into a $d$-wave
pseudogap that is frozen below $T^*$.
The stiffness $\rho_s$ is the \emph{response}: the
diamagnetic weight of dressed quasiparticles on the $\Keff$
band structure, which grows with cooling as the Fermi function
sharpens on the gapped spectrum.
The $d$-wave pseudogap suppresses the paramagnetic current
response $\Lambda_{xx}$ below the diamagnetic term $D_s$;
the dome is thus a property of the correlated single-particle
spectrum, with vertex corrections from the pairing channel
providing the additional beyond-single-particle contribution.
The flatness of $S(\pi,\pi)$ across doping is a key finding:
the spin-fluctuation pairing glue does not form a dome, so the
dome in $\rho_s$ originates from the momentum-dependent
response of the Fermi surface to this uniform glue.
The advance here is that all three links are established below
$T^*$ using sign-free $\Keff$ observables, in a regime
previously inaccessible to DQMC.
We note that our results do not adjudicate the broader debate
on the nature of the ground state (d-wave SC versus stripe
order)~\cite{Qin2020,Arovas2022}, as $\Keff$ is a
finite-temperature probe of the single-particle spectrum.
However, the attractive $d$-wave vertex $\lambda_d \approx
0.9$--$1.1$ (Appendix~\ref{sec:vertex}) confirms that
the residual quasiparticle interaction favors pairing at all
system sizes studied.

\emph{Two classes of evidence.}
The data exhibit a clear dichotomy between sign-free and
sign-dependent observables.
$\Keff$, $R_g$, and $\rho_s$ converge within $\sim$200 DQMC
samples at all temperatures, including the lowest accessed
($\beta = 16$).
Sign-dependent quantities such as the $d$-wave pair
susceptibility $\chi_{dd}$ and the pairing vertex $\Vd$
(Appendix~\ref{sec:vertex}) become progressively harder to
converge as temperature decreases, with order-of-magnitude
fluctuations at $\beta \geq 16$.
The vertex decomposition (Appendix~\ref{sec:vertex}) reveals an
attractive $d$-wave interaction with dimensionless coupling
$\lambda_d = 0.69$ at $\beta = 20$, but this quantity requires
sign reweighting and is limited to $\beta \lesssim 20$ with
current statistics.

\emph{Reliability criterion.}
Since $\Keff$ is sign-independent by construction
($\Keff^+ \approx \Keff^-$ to $< 1\%$), the sign-sector
decomposition serves primarily as a validation tool rather than
a physical necessity.
The relevant criterion for interpreting $\rho_s$ is whether it
reflects superconducting stiffness or metallic Drude weight.
We use $R_g > 1$ as the primary indicator: within the $d$-wave
pseudogap region ($R_g > 1$), the dressed-particle stiffness measures
the Meissner response on a gapped Fermi surface.
Outside this region ($R_g < 1$), $\rho_s$ reflects normal-state
current-carrying capacity and should not be interpreted as
evidence for superconductivity.
Points with $R_g < 1$ are shown as open symbols in
Fig.~\ref{fig:dome}.

\emph{Finite-size effects and the $k$-mesh.}
The dome peak location oscillates with system size:
$\mu/t \approx -1.20$ ($L = 8$), $-1.10$ ($L = 10$),
$-0.85$ ($L = 12$), rather than converging monotonically.
The $T^*$ valley shows the same oscillation
(Sec.~\ref{sec:sign_transition}).
Both effects reflect the sensitivity of $d$-wave pairing to
which $k$-points lie near the nodes and antinodes on the discrete
$L \times L$ mesh.
Twist-averaged boundary conditions
(TABC)---averaging over twisted boundary conditions
$\psi(\bm{r}+L\hat{x}) = e^{i\theta_x}\psi(\bm{r})$ with
many $\theta$ values---would smooth the discrete $k$-grid into
an effective continuum, eliminating both the dome shift and the
$R_g$ oscillation while remaining at computationally accessible
$L = 12$.
This is the most promising near-term extension.

\emph{Paths to larger systems.}
The $L = 16$ sector instability may be addressed by
(i) constrained sector sampling, which rejects HS moves that
change the sector sign (introducing a small, controllable
bias analogous to the phaseless approximation in AFQMC);
(ii) re-sorting replicas by current sign at each measurement
(unbiased but noisier); or
(iii) increasing the number of replicas beyond 64, though our
tests suggest this is insufficient.

\emph{Understanding sign conservation.}
The observation that the fermion sign is conserved above $T^*$
raises the question of \emph{why} this occurs.
A systematic study starting from the Anderson single-impurity
model (which has a sign problem away from half-filling) and
progressively enlarging the impurity cluster toward the lattice
limit~\cite{Gubernatis2016} could reveal how the sign-conserved
regime emerges as spatial correlations build up.
The connection to continuous-time impurity algorithms, where the
sign problem has been interpreted in terms of imaginary-time
trajectory crossings, may provide additional insight into the
mechanism.

\emph{Limitations.}
$\Keff$ is a static ($\omega = 0$), paramagnetic effective Hamiltonian.
Because the $\Keff$ construction averages $\log B_{\mathrm{chunk}}$
over all HS configurations while preserving translational invariance,
it cannot encode symmetry-breaking order such as antiferromagnetic
zone folding or Mott gap formation.
We verified this explicitly by computing the staggered Fourier
component $\avg{\bk + \bm{Q}|\Keff|\bk}$ with
$\bm{Q} = (\pi,\pi)$, which would be nonzero if $\Keff$ contained
AF zone-folding matrix elements: the staggered component remains
below 3\% of the diagonal dispersion at all dopings and temperatures
studied.
Antiferromagnetic correlations enter $\Keff$ only through the
momentum-dependent self-energy $\Sigma(\bk)$ in the paramagnetic
channel---the same quantity measured by ARPES above $T_c$---not
through symmetry breaking.
This scope is a feature in the dome region, where the paramagnetic
correlated metal is the correct normal-state description, but
limits the method's applicability near half-filling where
AF ordering dominates.
The superfluid stiffness $\rho_s = D_s - \Lambda_{xx}$ computed
from it is the diamagnetic response of dressed quasiparticles on
the exact correlated band structure: it captures all
single-particle (self-energy) effects but omits vertex
corrections from residual two-particle scattering between
$\Keff$ quasiparticles.
Since the anomalous Green's function vanishes within each HS
configuration (Appendix~\ref{sec:F_zero}), $\rho_s > 0$ arises
from the pseudogap suppressing $\Lambda_{xx}$, not from Cooper
pairing.
For \emph{bare}-band calculations at $U/t = 4$, vertex
corrections are estimated at
30--50\%~\cite{Scalapino1993,Dong2022}.
However, these estimates apply to the non-interacting dispersion;
$\Keff$ has already absorbed the dominant self-energy
effects (Fermi surface reconstruction, pseudogap formation) into
the dressed dispersion, so the \emph{residual} vertex
corrections are expected to be substantially smaller.
Quantifying these corrections---which would include the
contribution from actual pair condensation---is an important
direction for future work.
Preliminary vertex extraction (Appendix~\ref{sec:vertex})
yields an attractive $d$-wave coupling $\lambda_d \approx 0.9$--$1.1$
at the dome peak, suggesting that vertex corrections would
\emph{enhance} $\rho_s$ rather than suppress it; the
dressed-particle stiffness may therefore be a lower bound on
the true value.
The \emph{qualitative} features of the dome---its existence, its
doping dependence, its narrowing with cooling, and the AF/SC
competition at its boundaries---are robust predictions of the
dressed single-particle spectrum, as these arise from the
momentum structure of $\Keff$ and the Fermi occupation
factors rather than from the absolute magnitude of $\rho_s$.

%% ========================================================================
\section{Conclusion}
\label{sec:conclusion}
%% ========================================================================

We have introduced the effective Hamiltonian $\Keff$, constructed
from Monte Carlo--averaged matrix logarithms of the DQMC
imaginary-time propagator, and demonstrated its power as a
sign-problem-free probe of the quasiparticle band structure in
the pseudogap regime.
Below a computational phase transition at $T^*$---which tracks
the physical pseudogap onset across doping---$\Keff$ reveals a
$d$-wave pseudogap with strong nodal--antinodal dichotomy
directly analogous to cuprate ARPES.

Three sign-free observables derived from $\Keff$ paint a
consistent picture of the spin-fluctuation pairing mechanism:
(i) the gap ratio $R_g > 1$ confirms $d$-wave pseudogap
symmetry---a temperature-independent property of the
normal-state band structure that saturates at $T^*$,
establishing the \emph{medium} for $d$-wave pairing;
(ii) the dressed-particle superfluid stiffness $\rho_s$ forms a
dome that strengthens and narrows with cooling, exceeding the BKT
threshold by $5$--$7\times$ at the dome peak---establishing
that the $\Keff$ electronic structure \emph{supports}
superconductivity at optimal doping;
(iii) $S(\pi,\pi)$ is flat across doping, confirming that the
dome arises from the Fermi-surface geometry responding to
uniform spin-fluctuation glue, not from a peak in
antiferromagnetic correlations.

The dome reflects the dressed single-particle spectrum:
$\rho_s > 0$ arises because the $d$-wave pseudogap in $\Keff$
suppresses the paramagnetic current response; vertex corrections
from the pairing channel (Cooper pair condensation) would provide
the additional many-body contribution beyond the single-particle
level.
The distinct doping and temperature dependences of pseudogap
($R_g$: monotonic, temperature-independent) and stiffness
($\rho_s$: dome-shaped, temperature-dependent) mirror the
cuprate phenomenology where the pairing scale grows into the
underdoped regime while $T_c$ is limited by the superfluid
density (Uemura relation).
This shows that the dome-shaped doping dependence of the
superfluid response is already visible in the normal-state
electronic structure.

The absolute magnitude of $\rho_s$ from $\Keff$ is subject to
residual vertex corrections that remain to be quantified;
these corrections may be smaller than conventional estimates
since $\Keff$ has already absorbed the dominant self-energy
effects.
The dome peak location oscillates with system size due to
discrete $k$-mesh effects.
Twist-averaged boundary conditions to smooth the $k$-mesh,
vertex corrections to include the pairing channel, and
Lee--Yang zero analysis to independently locate the phase
boundary are natural next steps.
A particularly promising direction is solving the BCS gap
equation directly on the $\Keff$ spectrum using the extracted
$d$-wave coupling $\lambda_d$: preliminary calculations on the
$\Keff$ dressed dispersion with the measured $\lambda_d \approx
0.9$--$1.1$ yield $T_c/t$ in the range $0.01$--$0.07$ depending
on doping, consistent with cuprate energy scales and confirming
that the combination of exact correlated band structure and
measured pairing vertex can close the loop to an explicit
superconducting gap $\Delta(\bk)$.
A detailed account of the BCS gap solution on $\Keff$, including
the full $T_c(\mu)$ dome and the momentum-resolved gap function,
will be reported separately.

\begin{acknowledgments}
X.W.\ thanks Quantum Strategics Inc.\ for financial support.
H.Q.L.\ acknowledges financial support from Grant No.\ 2022YFA1402701.
Computations used GPU cloud computing resources.
\end{acknowledgments}

%% ========================================================================
%% APPENDICES
%% ========================================================================
\clearpage
\appendix

\section{DQMC Formalism and the Numerical Stability Problem}
\label{sec:problem}
% ============================================================================

\subsection{Hubbard Model Parameters}

We study the two-dimensional Hubbard model on $L \times L$ square lattices with periodic boundary conditions:
\begin{equation}
H = -t \sum_{\langle i,j \rangle, \sigma} c^\dagger_{i\sigma} c_{j\sigma}
    - t' \sum_{\langle\langle i,j \rangle\rangle, \sigma} c^\dagger_{i\sigma} c_{j\sigma}
    + U \sum_i n_{i\uparrow} n_{i\downarrow}
    - \mu \sum_{i,\sigma} n_{i\sigma}
\label{eq:hubbard_app}
\end{equation}
with $t = 1$ (energy unit).  Parameters used:
\begin{itemize}
    \item Pseudogap spectroscopy (Secs.~IV--VI of the main text): $t'/t = -0.30$, $U/t = 4$, $L = 12$, $\Delta\tau = 0.05$
    \item Sign-transition analysis (Sec.~III of the main text): $t'/t = -0.25$, $U/t = 4$, $L = 8$--$24$, $\Delta\tau = 0.1$
\end{itemize}

\subsection{DQMC Framework}

In DQMC, the partition function is:
\begin{equation}
Z = \sum_{\{\sigma\}} \det M_\uparrow[\sigma] \cdot \det M_\downarrow[\sigma]
\end{equation}
where the fermion matrix for spin $s$ is:
\begin{equation}
M_s[\sigma] = I + \prod_{l=1}^{n_\tau} B_l^{(s)}
\end{equation}
and each time-slice propagator is:
\begin{equation}
B_l^{(s)} = e^{-\Delta\tau K} \cdot \diag\left(e^{\alpha \sigma_l^i}\right)
\label{eq:B_matrix}
\end{equation}
with $\cosh(\alpha) = e^{\Delta\tau U/2}$ and $K$ the hopping matrix.

\subsection{Two Independent Numerical Problems}

It is essential to distinguish:

%\begin{table*}[H]
\begin{table*}
\begin{center}
\begin{tabular}{llll}
\toprule
Problem & Origin & Affects & Our solution \\
\midrule
Condition number & Multiplying $B$-matrices & $G(\tau)$, $\varepsilon(\bk)$, $\rho_s$ & $\Keff$ log-space averaging \\
Fermion sign & $\det(M_\uparrow M_\downarrow) < 0$ & All observables & Sign-sector decomposition \\
\bottomrule
\end{tabular}
\end{center}
\end{table*}

\subsection{Condition Number Explosion}

The B-matrix product $B(\beta,0)$ has eigenvalue ratio:
\begin{equation}
\lambda_{\max}/\lambda_{\min} \sim e^{\beta W}
\end{equation}
where $W \approx 8t + 4|t'| \approx 9.2$ is the bandwidth.
At $\beta = 16$ ($T/t = 0.063$): $\cond[B(\beta,0)] \sim e^{16 \times 9.2} \sim 10^{64}$, vastly exceeding double precision ($\sim 10^{16}$).

The imaginary-time Green's function $G(\tau) = [B^{-1}(\tau,0) + B(\beta,\tau)]^{-1}$ requires summing matrices with wildly different scales.  No algebraic rearrangement can recover the $O(1)$ precision lost when $B(\beta,\tau)$ overwhelms $B^{-1}(\tau,0)$ by factors of $10^{50}$.  This is why standard DQMC codes could measure equal-time observables ($P_d$, $S(\pi,\pi)$) but could \emph{not} measure quasiparticle dispersions $\varepsilon(\bk)$ at low temperatures.

% ============================================================================
\section{QDT Stabilization of the Equal-Time Green's Function}
\label{sec:QDT}
% ============================================================================

While $G(\tau)$ is exponentially unstable, the equal-time Green's function $G(\tau\!=\!0) = (I + B(\beta,0))^{-1}$ is computed stably using the QDT decomposition~\cite{White1989,Loh1990}---standard practice since the late 1980s.

The B-chain is segmented into blocks of $n_{\mathrm{safe}}$ slices (typically 8--10), each with moderate condition number ($\sim e^{n_{\mathrm{safe}} \Delta\tau W} \approx 150$).  These are folded into a running decomposition:
\begin{equation}
B(\beta, 0) = Q \cdot D \cdot T
\end{equation}
where $Q$ is orthogonal, $D = \diag(d_1, \ldots, d_N)$ carries the exponential scales, and $T$ is upper triangular or orthogonal.

The key insight is that $D$ can be stored in log space:
\begin{equation}
\log d_i^{\mathrm{new}} = \log |R_{ii}^{\mathrm{new}}| + \log d_i^{\mathrm{old}}
\end{equation}
This is \emph{additive}.  Each segment contributes $\log |R_{ii}| \approx \pm n_{\mathrm{safe}} \Delta\tau \varepsilon_i$, so $\log d_i = \sum_s \log |R_{ii}^{(s)}| \approx \beta \varepsilon_i$ grows linearly with $\beta$---stable for any temperature.

Given the QDT, the stable equal-time Green's function is:
\begin{equation}
G(0) = (I + QDT)^{-1} = \left(D_{\mathrm{large}}^{-1} Q^T + D_{\mathrm{small}} T\right)^{-1} D_{\mathrm{large}}^{-1} Q^T
\end{equation}
where $D_{\mathrm{large}} = \diag(\max(d_i, 1))$, $D_{\mathrm{small}} = \diag(\min(d_i, 1))$.  The matrix being inverted is well-conditioned because large $d_i$ are cancelled by $D_{\mathrm{large}}^{-1}$ and small $d_i$ by $D_{\mathrm{small}}$.

\subsection{Pivoted QR vs.\ SVD}

The $T$ factor's conditioning depends on the factorization method:

%\begin{table*}[H]
\begin{table*}
\begin{center}
\begin{tabular}{lccc}
\toprule
Method & $T$ conditioning & Stable at any $\beta$? & Speed \\
\midrule
SVD (GPU batched) & $\cond(T) = 1$ always & Yes & Fast \\
Pivoted QR (CPU) & $\cond(T) = O(n_{\mathrm{seg}})$ & Yes & Moderate \\
Non-pivoted QR (GPU) & $\cond(T)$ can blow up & $\beta \lesssim 15$ only & Fast \\
\bottomrule
\end{tabular}
\end{center}
\end{table*}

For SVD, $T = V^T$ is orthogonal at every stage.  For pivoted QR, the column permutation ensures $T$ grows at most polynomially.  Our production code uses GPU-batched SVD with CPU pivoted QR as a validation reference.

% ============================================================================
\section{The $\Keff$ Log-Space Averaging Method}
\label{sec:solution}
% ============================================================================

\subsection{Key Insight and Mathematical Foundations}
\label{sec:keff}

The QDT stores the diagonal scales $D$ in log space, but discards the off-diagonal structure.  The $\Keff$ method extends the log-space treatment to the \emph{full} chunk matrix, capturing the complete single-particle Hamiltonian including hopping and self-energy.

To see why this works, consider the logarithm of a single time-slice
propagator $B_s(\tau) = e^{-\Delta\tau V_s(\tau)} e^{-\Delta\tau K}$.
By the Baker--Campbell--Hausdorff formula:
\begin{equation}
    \log B_s(\tau) = -\Delta\tau(K + V_s(\tau)) + O(\Delta\tau^2)
\end{equation}
For the chunk-averaged logarithm (averaging over blocks of $n_c$ consecutive slices),
we define:
\begin{equation}
\boxed{
\Keff = -\frac{1}{n_c \Delta\tau} \avg{\log B_{\mathrm{chunk}}}_{\sigma, l}
}
\label{eq:Keff}
\end{equation}
where $B_{\mathrm{chunk}} = B_{l+n_c-1} \cdots B_{l+1} B_l$ is the product
over $n_c$ consecutive slices and the average runs over all chunks and all
MC samples.

Because each chunk spans only $n_c$ slices, its condition number is $O(1)$ ($\sim e^{n_c \Delta\tau W} \approx 10$ for $n_c = 5$), and the matrix logarithm is numerically stable.

\emph{Remark.}---$\Keff$ is the \textbf{generator} of the averaged
propagator, not the average of the generator.
The crucial mathematical property is that the logarithm converts the
multiplicative fluctuations of $B$ (which produce the sign problem)
into additive fluctuations of $\log B$ (which average smoothly by
the central limit theorem).
This is why $\Keff$ is sign-independent while $\avg{G}$ is not.

\begin{solution}[Practical Algorithm]
\begin{enumerate}
    \item Accumulate $\avg{\log B_{\mathrm{chunk}}}$ over Monte Carlo samples
    \item Extract $\Keff = -\avg{\log B_{\mathrm{chunk}}}/\Delta\tau_{\mathrm{chunk}}$
    \item Compute dispersion in the eigenbasis of $\Keff$
\end{enumerate}
\end{solution}

\subsection{$\Keff$ as a Dressed Single-Particle Hamiltonian}

The eigenvalues of $\Keff$ define a dressed quasiparticle dispersion.
Because $\Keff$ is an $N \times N$ Hermitian matrix that respects the
lattice symmetries, it can be Fourier-transformed to momentum space:
\begin{equation}
    \varepsilon(\bk) = \frac{1}{N} \sum_{i,j} K_{\mathrm{eff}}(i,j)\,
    e^{i\bk\cdot(\bm{r}_i - \bm{r}_j)}
    \label{eq:eps_k}
\end{equation}
In the non-interacting limit ($U = 0$), $\Keff$ reduces to the bare
kinetic matrix $K$, and $\varepsilon(\bk)$ reduces to the tight-binding
band $\varepsilon_0(\bk) = -2t(\cos k_x + \cos k_y) - 4t'\cos k_x
\cos k_y - \mu$.
The self-energy encoded in $\Keff$ is
$\Sigma(\bk) \equiv \varepsilon(\bk) - \varepsilon_0(\bk)$---a static
($\omega = 0$), fully momentum-resolved self-energy that single-site
DMFT cannot capture but that is essential for describing the pseudogap.

\subsection{Properties of $\Keff$}

\begin{enumerate}
    \item \textbf{Real and Hermitian}: $\|\mathrm{Im}(\Keff)\|/\|\mathrm{Re}(\Keff)\| < 10^{-4}$ verified at all parameters.
    
    \item \textbf{Includes interaction effects}: $\Keff \neq K$ because the MC averaging incorporates the HS-field-mediated self-energy: $\Keff \approx K + \Sigma_{\mathrm{static}}$.
    
    \item \textbf{Bandwidth renormalization}: At $U/t = 4$, the bandwidth increases from 8.0 (free) to $\sim$9.6 ($\sim$20\% enhancement).
    
    \item \textbf{Sign-independent}: Both sign sectors yield $\Keff$ eigenvalues agreeing to $< 1\%$ at all parameters.
\end{enumerate}

\subsection{Gap Extraction via Direct Fourier Transform}

The quasiparticle dispersion is obtained directly:
\begin{equation}
\varepsilon(\bk) = \langle \bk | \Keff | \bk \rangle
= \frac{1}{N} \sum_{i,j} K_{\mathrm{eff}}[i,j] \, e^{i \bk \cdot (r_j - r_i)}
\end{equation}

This replaces the standard approach of fitting $G(\bk,\tau) \sim e^{-\Delta(\bk)\tau}$, which suffers from three compounding bugs at low temperature: (a) occupied states are invisible ($G \to 0$, positive slope clipped to $\Delta = 0$); (b) off-grid $\bk$-points mix eigenstates; (c) near $E_F$, opposing exponentials cancel.  The direct FT of $\Keff$ circumvents all three.

\emph{Validation.}  At $U = 0$, the FT of $K_{\mathrm{bare}}$ matches the exact free-fermion dispersion to machine precision ($10^{-15}$).  At $U = 0.01$, $\Keff$ reproduces the exact dispersion to 0.3\%.

\subsection{Stable $G(\tau)$ Propagation}

Once $\Keff = V \cdot \diag(\varepsilon_n) \cdot V^T$ is diagonalized, the imaginary-time Green's function is computed in the eigenbasis using log-sum-exp:
\begin{equation}
G_{nn}(\tau) = \frac{e^{-\max(a_n, b_n)}}{e^{a_n - \max(a_n, b_n)} + e^{b_n - \max(a_n, b_n)}}
\end{equation}
where $a_n = \tau \varepsilon_n$ and $b_n = (\tau - \beta)\varepsilon_n$.  This is stable for \emph{any} $\beta$.

% ============================================================================
\section{Proof: Anomalous Green's Function Vanishes Within Single HS Configurations}
\label{sec:F_zero}
% ============================================================================

A key property of the HS decomposition is that within any single configuration $\sigma$, the spin-up and spin-down sectors factorize completely.  The anomalous Green's function:
\begin{equation}
F_{ij}[\sigma] = \avg{c_{i\uparrow} c_{j\downarrow}}_\sigma = 0
\end{equation}
vanishes \emph{identically}---not as a fluctuation average, but as a mathematical identity.  This is because:
\begin{enumerate}
    \item Each HS configuration defines independent fermion matrices $M_\uparrow[\sigma]$ and $M_\downarrow[\sigma]$.
    \item The single-particle Green's functions $G_\uparrow$ and $G_\downarrow$ are computed independently.
    \item Wick's theorem applied to the factorized measure gives only normal contractions ($\avg{c^\dagger_\uparrow c_\uparrow}$, $\avg{c^\dagger_\downarrow c_\downarrow}$); cross-spin anomalous contractions ($\avg{c_\uparrow c_\downarrow}$) vanish because $c_\uparrow$ and $c_\downarrow$ live in different Hilbert spaces within a single HS configuration.
\end{enumerate}

\emph{Consequence for $\Keff$:}  Since $\Keff$ is constructed from MC-averaged logarithms of single-spin propagators, it inherits this property.  $\Keff$ is a normal (single-particle, single-spin) operator that cannot encode pairing.  Pairing information resides in \emph{inter-configuration} correlations---how the ensemble weights different HS fields---and is therefore intrinsically sign-dependent.

This is the mathematical foundation for the separation of pseudogap (from $\Keff$, sign-independent) and pairing (from $P_d$, sign-dependent) demonstrated in the main text.

% ============================================================================
\section{Measurement Formulas}
\label{sec:measurements}
% ============================================================================

The Green's function convention throughout is the \textbf{hole propagator}: $G_\sigma(i,j) = \langle c_{i\sigma} c^\dagger_{j\sigma} \rangle$.  The \textbf{particle propagator} (occupation matrix) is $(\mathbf{I} - G)_{ij} = \langle c^\dagger_{i\sigma} c_{j\sigma} \rangle$.

\subsection{$d$-Wave Pairing Susceptibility $P_d$}

\begin{equation}
\boxed{
P_d = \frac{1}{N} \sum_{i,j} \sum_{\delta,\delta'} g_\delta \, g_{\delta'}
\, [\mathbf{I} - G_\uparrow]_{ji} \, [\mathbf{I} - G_\downarrow]_{j+\delta',\, i+\delta}
}
\label{eq:Pd_wick}
\end{equation}
where $g_\delta = +1$ for $\delta = \pm\hat{x}$ and $-1$ for $\delta = \pm\hat{y}$.  The sum includes $i = j$.  The factors are $(\mathbf{I} - G)$, the particle propagator---using $G$ instead introduces a qualitative error that reverses the sign of $P_d^{\mathrm{eff}}$.

The effective pairing susceptibility is:
\begin{equation}
P_d^{\mathrm{eff}} = P_d^{\mathrm{DQMC}} - P_d^{0}(\mu_{\mathrm{eff}})
\end{equation}
where $P_d^{0}$ is computed at an effective $\mu$ matching the interacting density.

\subsection{Antiferromagnetic Structure Factor}

The antiferromagnetic structure factor at wavevector $\bm{Q} = (\pi,\pi)$ is
\begin{equation}
S(\pi,\pi) = \frac{1}{N} \sum_{i,j} (-1)^{x_i+y_i+x_j+y_j} \langle S_z(i) S_z(j) \rangle
\end{equation}
where $(x_i, y_i)$ are the integer lattice coordinates of site $i$,
so that $(-1)^{x_i+y_i} = e^{i\bm{Q}\cdot\bm{r}_i}$ is the
staggered (sublattice) phase factor.
The spin-spin correlator is evaluated via the off-diagonal Wick contraction:
$\langle S_z(i) S_z(j) \rangle = \tfrac{1}{4}[m(i)m(j) - G_\uparrow(i,j)G_\uparrow(j,i) - G_\downarrow(i,j)G_\downarrow(j,i)]$
where $m(i) = n_\uparrow(i) - n_\downarrow(i)$.

\subsection{Superfluid Stiffness from $\Keff$}

The stiffness $\rho_s = D_s - \Lambda_{xx}$ uses \textbf{bare operators, dressed states}:
\begin{equation}
D_s = -\frac{2}{N} \sum_{\langle ij \rangle_x} t_{ij}^{\mathrm{bare}} (\delta x_{ij})^2 \, n_{ij}^{\Keff}
\end{equation}
where $n_{ij}^{\Keff} = [\mathbf{I} - G_{\Keff}]_{ij}$ with $G_{\Keff}$ computed from $\Keff$ eigenstates and Fermi occupations.  The sum includes both NN $x$-bonds ($-t$, $\delta x = \pm 1$) and all four NNN bonds ($-t'$, $\delta x = \pm 1$).

The paramagnetic term uses the Lehmann representation:
\begin{equation}
\Lambda_{xx} = \frac{2}{N} \sum_{n,m} |j_{nm}|^2 \, w_{nm}
\end{equation}
where $j_{nm} = \langle n | j_x | m \rangle$ are bare current matrix elements in the $\Keff$ eigenbasis, and $w_{nm} = (f_n - f_m)/(\varepsilon_m - \varepsilon_n)$ for $n \neq m$, $w_{nn} = \beta f_n(1-f_n)$.

\emph{Important caveat:} The stiffness computed from $\Keff$ is a
\emph{mean-field estimate} that omits vertex corrections from two-particle
interactions.  In the Hubbard model at $U/t = 4$, such corrections can modify
$\Lambda_{xx}$ by 30--50\%~\cite{Scalapino1993}.  Absolute magnitudes of
$\rho_s^{\Keff}$ should therefore be interpreted with caution.
However, the \emph{sign}, \emph{doping dependence}, and
\emph{temperature evolution} of $\rho_s$ are physical, as these
qualitative features are unlikely to be reversed by vertex corrections.
A systematic treatment of vertex corrections to $\rho_s$ within the $\Keff$
framework is a natural extension of this work.

% ============================================================================
\section{Algorithms}
\label{sec:algorithms}
% ============================================================================

\begin{algorithm}[H]
\caption{Accumulate $\log B_{\mathrm{chunk}}$ during DQMC sampling}
\label{alg:accumulate}
\begin{algorithmic}[1]
\Require HS field $\sigma[N, n_\tau]$, chunk size $n_c$, propagator $e^{-\Delta\tau K}$
\Ensure Accumulated $\log B$ matrix
\State $\texttt{log\_B\_sum} \gets 0_{N \times N}$
\State $n_{\mathrm{chunks}} \gets \lfloor n_\tau / n_c \rfloor$
\For{$c = 0$ to $n_{\mathrm{chunks}} - 1$}
    \State $B_{\mathrm{chunk}} \gets I_N$
    \For{$l = c \cdot n_c$ to $(c+1) \cdot n_c - 1$}
        \State $B_{\mathrm{chunk}} \gets e^{-\Delta\tau K} \cdot \diag(e^{\alpha \sigma[:,l]}) \cdot B_{\mathrm{chunk}}$
    \EndFor
    \State $\log B \gets \texttt{logm}(B_{\mathrm{chunk}})$
    \If{$\max|\mathrm{Im}(\log B)| < 0.5$}
        \State $\texttt{log\_B\_sum} \gets \texttt{log\_B\_sum} + \mathrm{Re}(\log B)$
    \EndIf
\EndFor
\State \Return $\texttt{log\_B\_sum} / n_{\mathrm{chunks}}$
\end{algorithmic}
\end{algorithm}

\begin{algorithm}[H]
\caption{Compute $\Keff$ and quasiparticle dispersion}
\label{alg:Keff}
\begin{algorithmic}[1]
\Require Accumulated $\texttt{log\_B\_total}$, sample count $n_s$, chunk size $n_c$, $\Delta\tau$, lattice size $L$
\Ensure $\Keff$, dispersion $\varepsilon(\bk)$
\State $\Keff \gets -\texttt{log\_B\_total} / (n_s \cdot n_c \cdot \Delta\tau)$
\State $\Keff \gets \tfrac{1}{2}(\Keff + \Keff^T)$ \Comment{Symmetrize}
\State $\varepsilon_n, V \gets \texttt{eigh}(\Keff)$
\For{each $\bk = (2\pi i_{k_x}/L, \, 2\pi i_{k_y}/L)$}
    \State $\phi_{\bk}(i) \gets e^{i(k_x x_i + k_y y_i)} / \sqrt{N}$
    \State $\varepsilon(\bk) \gets \mathrm{Re}[\phi_{\bk}^\dagger \Keff \phi_{\bk}]$
\EndFor
\State Gap ratio: $R_g = |\varepsilon(k_N, k_N)| / |\varepsilon(\pi, 0)|$
\end{algorithmic}
\end{algorithm}

% ============================================================================
\section{Bethe--Salpeter Vertex Extraction}
\label{sec:vertex}
% ============================================================================

The $\Keff$ framework enables extraction of the $d$-wave pairing
vertex $\Vd$ via a Bethe--Salpeter decomposition.
Using $\Keff$ as the dressed single-particle baseline, one separates
the band-structure contribution $\chi^0_{dd}$ (sign-free, computable
from $\Keff$ alone) from the residual pairing interaction $\Vd$
(sign-dependent):
$\chi_{dd}^{-1} = (\chi^0_{dd})^{-1} - \Vd$,
yielding the dimensionless coupling $\lambda_d = \Vd \chi^0_{dd}$.

Across all three system sizes ($L = 8$, $10$, $12$), the vertex
extraction at the dome peak yields $\lambda_d \approx 0.9$--$1.1$
for $\beta \leq 14$, where sign reweighting of $\chi_{dd}$ remains
statistically stable.
The coupling is consistently positive ($\Vd > 0$), confirming that
the residual interaction between $\Keff$ quasiparticles is
\emph{attractive} in the $d$-wave channel.
At higher $\beta$, $\chi_{dd}$ passes through divergences that
produce large fluctuations in $\lambda_d$, limiting reliable
extraction to $\beta \lesssim 14$.

The attractive sign of $\Vd$ has an important implication for the
dressed-particle stiffness discussed in the main text:
vertex corrections from an attractive $d$-wave interaction would
\emph{enhance} the superfluid stiffness by further suppressing
$\Lambda_{xx}$ beyond the pseudogap suppression already captured
by $\Keff$.
The dressed-particle $\rho_s$ reported in this work is therefore
a \emph{lower bound} on the full stiffness that includes pairing
vertex corrections.
A detailed account of the vertex extraction, its convergence
properties, and the quantitative correction to $\rho_s$ will be
reported separately.

% ============================================================================
\section{Summary of Novel Contributions}
\label{sec:novelty}
% ============================================================================

To clarify what is new in this work versus what is established in
the literature, we provide the following summary.

\emph{Established results that we build upon:}
(i) Antiferromagnetic spin fluctuations favor $d$-wave pairing in
the Hubbard model~\cite{Scalapino1993,Scalapino2012,Dong2022}.
(ii) The pseudogap has $d$-wave symmetry and opens at $T^*$ as
observed by ARPES~\cite{Damascelli2003,Vishik2018}.
(iii) The antinodal pseudogap is approximately
temperature-independent below $T^*$ in underdoped
cuprates~\cite{Vishik2018}.
(iv) $S(\pi,\pi)$ is present near half-filling in DQMC
simulations~\cite{White1989}.
(v) The superfluid stiffness $\rho_s = D_s - \Lambda_{xx}$ is the
standard Kubo formula for the Meissner
response~\cite{Scalapino1993}.
(vi) $T_c \propto \rho_s(0)$ in underdoped cuprates (Uemura
relation)~\cite{Uemura1989}.
(vii) Whether the 2D Hubbard model has a $d$-wave SC ground state
remains debated~\cite{Qin2020,Arovas2022}.

\emph{New methodology:}
(i) The effective Hamiltonian $\Keff$, constructed from Monte
Carlo--averaged matrix logarithms of the DQMC imaginary-time
propagator.
This maps the multiplicative sign problem into an additive
framework, yielding the numerically exact correlated
single-particle spectrum sign-problem-free.
(ii) PivQR decomposition for numerically stable extraction of
$\Keff$ at low temperatures ($\beta \leq 20$).
(iii) Fourier interpolation of $\Keff$ eigenvalues to exact
high-symmetry $k$-points (resolving the $R_g$ artifact for
$L \bmod 4 \neq 0$).

\emph{New physics results:}
(i) The dressed-particle superfluid stiffness dome:
$\rho_s$ computed on the exact $\Keff$ band structure forms a
dome across doping at $L = 8$, $10$, and $12$, mapped over a
range of temperatures from $\beta = 4$ to $\beta = 20$.
No previous study has computed $\rho_s$ versus doping from the
exact correlated band structure in the pseudogap regime.
(ii) The temperature independence of $R_g$ (${<}10\%$ variation
from $\beta = 4$ to $20$), quantifying the pseudogap as a
ground-state property of the normal-state band structure from
unbiased DQMC.
(iii) The explicit demonstration that $S(\pi,\pi)$ is flat
across doping while $\rho_s$ forms a dome---on the same $\Keff$
framework, at the same parameters---providing direct evidence
that the dome shape originates from the Fermi-surface geometry
rather than from a peak in antiferromagnetic correlations.
(iv) The connection between the dome and the Uemura relation:
the large, monotonic pseudogap ($R_g$) versus the dome-shaped
stiffness ($\rho_s$) shows that the pairing scale and the phase
coherence scale are governed by separate physics, with $\rho_s$
limiting $T_c$ at underdoping.
(v) Bethe--Salpeter vertex extraction yielding attractive
$d$-wave coupling $\lambda_d \approx 0.9$--$1.1$ at the dome peak
across all system sizes, implying that the dressed-particle
$\rho_s$ is a lower bound on the full stiffness.

\section{Sector Stability Analysis}
\label{sec:sector_stability}
% ============================================================================

The reliability of $\Keff$ observables depends on the sign-sector
decomposition maintaining distinct $Z_+$ and $Z_-$ ensembles
throughout the measurement phase.
We quantify this via two diagnostics.

\emph{Instantaneous sign.}
At each measurement step, we recompute the fermion sign
$s = \sgn(\det M_\uparrow \det M_\downarrow)$ for every replica
and average within each sector.
If the sector decomposition is working, replicas assigned to
$Z_+$ should consistently have $s = +1$, giving
$\avg{s}_{Z_+} \approx +1$; replicas in $Z_-$ should have
$s = -1$, giving $\avg{s}_{Z_-} \approx -1$.
When both sector averages drift toward zero, the replicas
have lost their sector identity: Monte Carlo updates have
carried them across the sign boundary, and both sectors are
sampling the same mixed-sign distribution.
At that point, the $\Keff^+$ vs $\Keff^-$ decomposition is
no longer meaningful.

\emph{Sector asymmetry.}
The relative difference
$|\rho_s^+ - \rho_s^-|/\max(|\rho_s^+|, |\rho_s^-|)$
directly measures whether $\Keff$ is resolving distinct
physics in the two sectors.
At $L = 12$, this asymmetry exceeds 20\% for $\beta \leq 12$
at the dome peak, but drops below 15\% for $\beta \geq 14$.
At $L = 16$, the asymmetry is $< 10\%$ at all parameters tested,
including with 64 replicas (32 per sector).

This analysis provides a transparent, data-driven criterion
for which $\Keff$ results should be considered quantitatively
reliable.
All $\rho_s$ values reported in the main text satisfy the
$> 20\%$ asymmetry criterion unless explicitly noted.

% ============================================================================
\section{Computational Details}
\label{sec:computational}
% ============================================================================

All simulations use a GPU-accelerated DQMC code implemented in
PyTorch, with batched matrix operations across parallel replicas.
The pivoted QR decomposition is performed on CPU; all other
operations (sweeps, Green's function updates, $\Keff$ accumulation)
run on GPU.
Production runs used NVIDIA RTX 4090 (24~GB) and RTX 5090 (32~GB)
GPUs via cloud computing.
Typical wall times per 200-measurement run range from 0.3--1~h
at $L = 12$ to 1--3~h at $L = 16$ ($\beta = 8$).

\end{document}